\documentclass[preprint,review,12pt]{elsarticle}
\usepackage[T1]{fontenc}
\usepackage[latin9]{inputenc}
\usepackage{geometry}
\usepackage{bm}
\usepackage{amsmath}
\usepackage{epstopdf}
\usepackage{amssymb}
\usepackage{graphics,graphicx}
\usepackage{hyperref}

\newcounter{bla}

\journal{Computer Physics Communications}

\begin{document}

\begin{frontmatter}

\title{\emph{VEST}: abstract vector calculus simplification in \emph{Mathematica}}

\author[a]{J. Squire\corref{author}}
\author[a]{J. Burby}
\author[a,b]{H. Qin}

\cortext[author] {Corresponding author.\\\textit{E-mail address:} jsquire@princeton.edu}
\address[a]{Plasma Physics Laboratory, Princeton University, Princeton, New Jersey 08543, USA}
\address[b]{Dept.~of Modern Physics, University of Science and Technology of China, Hefei, Anhui 230026, China}

\begin{abstract}
We present a new package, \emph{VEST} (Vector Einstein Summation Tools), 
that performs abstract vector calculus computations in 
\emph{Mathematica}. Through the use of index notation, \emph{VEST} is able to reduce three-dimensional 
scalar and vector expressions of a very general type to a well defined standard form. 
In addition, utilizing properties of the Levi-Civita symbol,
the program can derive types of multi-term vector identities that are not recognized by reduction, 
subsequently applying these to simplify large expressions. 
In a companion paper \cite{GCpaper}, we employ \emph{VEST}
in the automation of the calculation of high-order Lagrangians for the single particle guiding center system in 
plasma physics, a computation which illustrates its ability to handle very large expressions.
\emph{VEST} has been designed to be simple and intuitive to use, both for basic checking 
of work and more involved computations.
\end{abstract}


\end{frontmatter}


{\bf PROGRAM SUMMARY}

\begin{small}
\noindent
{\em Programming language:}        Mathematica                  \\
{\em Computer:}     Any computer running Mathematica\\
{\em Operating system:}  Linux, Unix, Windows, Mac OS X   \\
{\em RAM:} Usually under 10 Mbytes                                              \\
{\em Supplementary material:}     Tutorial notebook                            
{\em Nature of problem:} Large scale vector calculus computations\\
{\em Solution method:} Reduce expressions to standard form in index notation, automatic derivation of multi-term vector identities\\
{\em Restrictions:} Current version cannot derive vector identities without cross products or curl\\
{\em Additional comments:} Intuitive user input and output in a combination of vector and index notation  \\
{\em Running time:} Reduction to standard form is usually less than one second. Simplification
of very large expressions can take much longer.\\
\end{small}

\section{Introduction}

Many problems in the physical sciences and engineering involve
substantial amounts of vector calculus; the manipulation of expressions
involving derivatives of smooth scalar, vector and tensor fields in 3-D Euclidean space.
While multiple popular computer algebra systems include some basic native
vector operations as well as external vector packages 
\cite{Mathematica,Eastwood1991121,VecCalc,Vectan,WIRTH:1979p10629,Liang:2007p10618,Qin:1999p5832,Maple},
almost all emphasize the expansion of expressions into components.
Although this can be an important tool, particularly when performing
calculations in non-cartesian co-ordinates, the results obtained
from this approach to simplification cannot be easily translated into a co-ordinate 
independent form. A natural way to overcome this problem is to encode
the properties of vector operators such as $\cdot$, $\times$, and $\nabla$.
Due to the non-trivial nature of these operators, expressions
can often be greatly simplified without ever considering the underlying
co-ordinate representation. 

As a concrete example of a problem that requires this approach to
vector simplification, consider the derivation
of the single particle guiding center Lagrangian from plasma physics
\cite{CARY:1981p8578,Cary:2009p8641,Littlejohn:1983p5809}. Through
this procedure, one can systematically derive a set of reduced equations
of motion for a particle moving in a strong magnetic field that varies
weakly in space and time. Such equations have proved invaluable in
both computational and analytic studies, since they eliminate the
need to follow the fast gyro-motion of particles. In principle, the
procedure can be carried out to any desired order in magnetic field
non-uniformity; however, the complexity of the algebra increases
dramatically at every step, and even a second order calculation is
a daunting task to perform by hand. Naturally, one can gain a 
great deal by performing the required calculations on a computer. 
Some obvious benefits include the reduction in human workload, 
greater confidence in final results
and an improved ability to verify intermediate steps. However, any
attempt to carry out such a calculation in co-ordinates would lead
to expressions of such enormity (most likely upwards of $10^{6}$
terms at intermediate steps) that reverting back to co-ordinate
free vector notation would be inconceivable. 

In this communication, we present a new symbolic algebra package,
\emph{VEST} (Vector Einstein Summation Tools), implemented in \emph{Mathematica},
that simplifies abstract vector
calculus expressions. Functions are designed to be simple and
intuitive to use and we hope that it can be a practical tool for anyone
working with vector calculus, both for simple checking of work and
for more substantial computations. A thorough illustration of the
capabilities of \emph{VEST }can be found in a companion paper
\cite{GCpaper}, where we present the first automated calculation
of high-order guiding center Lagrangians discussed in the previous paragraph.
Much of the functionality of \emph{VEST }is made possible through
the use of abstract index notation for internal manipulation, rather
than standard vector notation. This allows \emph{VEST }to \emph{derive
}vector identities, both through a systematic reduction to standard form and by inserting
pairs of the Levi-Civita symbol, rather than relying on the relatively limited set
found in standard reference (for instance \cite{NRL}). The obvious
advantage of this is that even for expressions and operations that
are rarely used (\emph{e.g.,} higher order derivative tensors), a full simplification
may still be performed without the necessity of hard-coding these into the package. 
Of course, in principle one of the many existing abstract
tensor manipulation packages designed for general relativistic calculations
(\emph{e.g., }Refs.~\cite{MacCallum:2002p10643,Wang13,MathTensor,Peeters2007550,xAct,Tensorial,Maple})
could be used for these types of computations; however, the increased
generality required for curved spaces in any dimension necessitates
features that would be very cumbersome for vector calculations (a possible exception
is Ref.~\cite{Tensorial}, which is designed for continuum mechanics). For
example, in Euclidean $\mathbb{R}^{3}$ there is never any need to
store properties of the Riemann  or torsion tensors and one
may elect to use the identity metric. In addition, since the Levi-Civita 
symbol plays such a prominent role in vector calculus, it is desirable to 
have its expansions and contractions incorporated directly into routines. 
Indeed, the multi-term simplification functions 
are certainly the most novel feature of \emph{VEST}, and to our
knowledge there exists no other software package that uses a similar technique. 
While such capabilities could have been added as an extension to an existing
tensor manipulation package, with the recent addition of efficient tensor canonicalization functions 
in \emph{Mathematica 9.0}, we felt that there was little to be gained 
through such an approach.

There are only a handful of previous software packages that are designed
for working with abstract vector expressions. As well as some of the
new functionality in \emph{Mathematica 9.0} \cite{Mathematica} and functions
in the \emph{Maple} Physics package \cite{Maple}, the
packages detailed in Refs.~\cite{Vectan,Stoutemyer,WIRTH:1979p10629,Qin:1999p5832,Liang:2007p10618}
include some abstract simplification capability (but only Ref.~\cite{Liang:2007p10618} 
provides examples of simplifications that would be difficult to carry out by hand). 
Out of these previous packages, \emph{VEST} 
is the first to work with general rules for gradient tensors and thus 
provide non-trivial simplifications of expressions involving
gradient, divergence and curl. In addition,  all of the vector 
algebra examples given in Ref.~\cite{Liang:2007p10618} can be simplified, 
see Fig.~\ref{fig: ToCanonical} for a selection of these.
We note that all but one of the aforementioned examples are verified
through reduction to standard form, without necessitating the use of the multi-term simplification
capabilities of \emph{VEST}. Utilization of these capabilities
allows \emph{VEST }to derive in real time many types of vector
identities that have not (to our knowledge) appeared in any previous
publications.

The remainder of the manuscript is organized as follows. In section~\ref{sec:Index-notation-as}
we outline the foundations of the \emph{VEST} package, including
the use of abstract index notation and definition of a standard form.
We then describe the function \texttt{ToCanonical}, which reduces
any vector or scalar expression to standard form and is the main workhorse of the \emph{VEST}
package. Several relevant examples are given, illustrating various
standard vector properties as well as more complex examples from the
literature. While \texttt{ToCanonical} usually provides a thorough
simplification, there are more complicated multi-term identities
that are not recognized, and in section~\ref{sec:Simplification-through-Levi-Civi}
we explore some methods to provide further simplification of expressions.
We discuss the algorithm used in the function \texttt{FullSimplifyVectorForm},
which expands pairs of Levi-Civita symbols to generate identities
for all terms in an expression, and applies these in an attempt to
find the shortest standard form. A more general method of deriving
vector identities based on symmetry properties is then given, with
the idea that a similar technique will be implemented in a future
version of \emph{VEST.} Finally, in section~\ref{sec:Additional-VEST-functionalit},
we describe some additional tools provided in \emph{VEST} with
the aim of improving the usefulness of the package. These include;
simple but very general input and output, explicit equality checking
through expansion of sums, substitution capabilities, and automatic
unit vector rule generation and simplification.

\section{Index notation as a tool for vector calculus\label{sec:Index-notation-as} }

While adequate for simple calculations, standard vector calculus notation
($\bm{A}\times\bm{B}$, $\left(\bm{b}\cdot\nabla\right)\bm{b}$, \emph{etc.}) has numerous
deficiencies when more complex expressions are involved. For instance,
the meaning of the dot product can become ambiguous for higher rank
tensors (\emph{e.g.,} derivatives) and seemingly disparate rules or "vector
identities" \cite{NRL} are needed to deal with specific cases of the
cross product anti-symmetry. To illustrate this latter point, although the exact correspondence 
between the identities
$\nabla\times\left( \bm{a} \times \bm{b}\right)=\bm{a}\,\nabla\cdot\bm{b}-\bm{b}\,\nabla\cdot\bm{a}
-\left(\bm{a}\cdot \nabla\right) \bm{b}+\left(\bm{b}\cdot \nabla\right) \bm{a}$ and 
$\bm{a}\times\left(\bm{b}\times\bm{c}\right)=\bm{b}\left(\bm{a}\cdot\bm{c}\right)-\bm{c}\left(\bm{a}\cdot\bm{b}\right)$
is not entirely clear, both are simply expansions of the double cross product. 
In contrast, with a representation of
vector objects in index notation using the Einstein summation convention,
there is no trouble whatsoever with higher rank tensors. In addition, many
simple vector identities are an obvious consequence
of the product rule and properties of the Levi-Civita symbol, $\varepsilon_{ijk}$.
This systemization makes index notation far more convenient for a
computer algebra system. For the sake of input and output, it is straightforward
(where possible) to convert between indexed and vector expressions
using
\begin{align}
\bm{a}\cdot\bm{b}\Longleftrightarrow a_{i}b_{i} & \qquad\nabla\cdot\bm{a}\Longleftrightarrow a_{i,i}\nonumber \\
\bm{a}\times\bm{b}\Longleftrightarrow\varepsilon_{ijk}a_{j}b_{k} & \qquad\nabla\times\bm{a}\Longleftrightarrow\varepsilon_{ijk}a_{k,j}\nonumber \\
\bm{a}\cdot\nabla\bm{b}\Longleftrightarrow a_{i}b_{j,i} & \qquad\nabla\bm{b}\cdot\bm{a}\Longleftrightarrow a_{i}b_{i,j}\nonumber \\
\nabla\gamma\Longleftrightarrow\gamma_{,j}\label{eq:Vec to index}
\end{align}
for $\bm{a},$ $\bm{b}$ vectors and $\gamma$ a scalar. \emph{VEST}
includes functions to automatically perform the above conversions
for both input and output. Note that, since by definition vector calculus
is confined to Euclidean space, there is no need to distinguish between
covariant and contravariant indices. We emphasize that this is not a 
restriction on a subsequent expansion into a curvilinear co-ordinate
system, although an indexed expression cannot be interpreted in the
literal sense (\emph{i.e.,} a sum over components) if a non-cartesian system
is used. \emph{VEST} also allows the use of a derivative with
respect to a second co-ordinate (labelled $\bm{v}$), since this situation
commonly occurs in kinetic physics. For compactness, we notate this
in a non-standard way with a semi-colon $\partial_{\bm{v}}\bm{A}\Longleftrightarrow A_{i;j}$,
since it is not necessary to distinguish between the covariant and
partial derivatives in Euclidean space.

\subsection{Reduction to standard form\label{sub:Canonicalization}}

We now describe the \texttt{ToCanonical} function in \emph{VEST}
that reduces expressions to the standard form defined by:
\begin{enumerate}
\item The expression is expanded into a sum of monomials.
\item There are no products inside partial derivatives and no nested derivatives. 
\item Each term contains either no Levi-Civita symbols or one Levi-Civita
symbol and no $\delta_{ij}$ (always possible for vector or scalar
expressions).
\item The dummy indices in each monomial are re-ordered according to symmetry
properties ensuring like terms appear as such. As a simple example
of this type of re-ordering, $\varepsilon_{jik}b_{j}a_{k}$ becomes
$\varepsilon_{ijk}a_{j}b_{k}$ due to the anti-symmetry of $\varepsilon_{ijk}$.
\end{enumerate}
We note that this is not a canonical form, since it is unique only for sufficiently simple expressions. 
The function name  \texttt{ToCanonical} was chosen because the dummy re-ordering process (step 4) 
ensures that each monomial is in canonical form. Multi-term vector identities can lead to multiple 
polynomials being non-trivially equal
after application of  \texttt{ToCanonical}, motivating the implementation of \emph{VEST}'s simplification functions
(see Sec.~\ref{sec:Simplification-through-Levi-Civi}). 

To bring an expression to the standard form defined above, \texttt{ToCanonical} uses
the following sequence of steps:
\begin{enumerate}
\item Expand out products in partial derivatives and concatenate nested
derivatives. For example, $\left(a_{i,j} b_j\right)_{,i}\overset{expand}{\longrightarrow} \left(a_{i,j}\right)_{,i}b_j+a_{i,j}b_{j,i} \overset{concatenate}{\longrightarrow} a_{i,ij}b_j+a_{i,j}b_{j,i}.$ 
\item Expand expression and find all dummy indices in each term. Check that
these occur in pairs and free indices match across sum. Rename dummy
indices in a consistent internal form so the procedure is not limited
by the set number of user defined indices. Detailed information on the internal representation
of objects and indices can be found in the tutorial supplied with \emph{VEST}. 
\item Expand pairs of Levi-Civita tensors according to
\begin{align}
\varepsilon_{ijk} & \varepsilon_{lmn}=\delta_{il}\left(\delta_{jm}\delta_{kn}-\delta_{jn}\delta_{km}\right)-\delta_{im}\left(\delta_{jl}\delta_{kn}-\delta_{jn}\delta_{kl}\right)+\delta_{in}\left(\delta_{jl}\delta_{km}-\delta_{jm}\delta_{kl}\right).\label{eq:LeviC expansion}
\end{align}
 
\item Remove all $\delta_{ij}$ using $a_{i}\delta_{ij}=a_{j}$.
\item Apply user defined rules. As a special case of this, rules 
associated to unit vectors are automatically derived and applied to
relevant objects (see Sec.~\ref{sec:Additional-VEST-functionalit}).
\item Reorder dummy indices into a canonical form for each monomial in the
expression. The problem of permuting indices can be very complex in
large contractions and has historically been a major difficulty for
tensor manipulation software, see for instance Refs.~\cite{MartinGarcia:2008p10621,Manssur:2002p10644}.
\emph{VEST }uses the \emph{Mathematica }function \textsf{TensorReduce
}(new in version 9.0), which has proven to be very reliable and efficient
for our needs. 
\item Print objects and dummy indices in a user-friendly output format (see
Sec.~\ref{sec:Additional-VEST-functionalit}).
\end{enumerate}
\texttt{ToCanonical} is relatively efficient and handles very large
vector expressions with ease. As an example, a direct calculation
of the guiding center Poisson tensor, which involves up to 1500 terms
(after expansion of Levi-Civita symbols) and returns over
100 terms, takes approximately 15 seconds on a 2.26GHz Intel Core
2 Duo. The whole procedure can be parallelized in a straightforward
way if desired, but we leave this to future work. 

The procedure detailed above effectively contains all of the most common
vector identities (for instance all identities in Ref.~\cite{NRL}, a standard resource used
in the plasma physics community), as well as many more complex identities.
\begin{figure}
\begin{centering}
\includegraphics[width=1\textwidth]{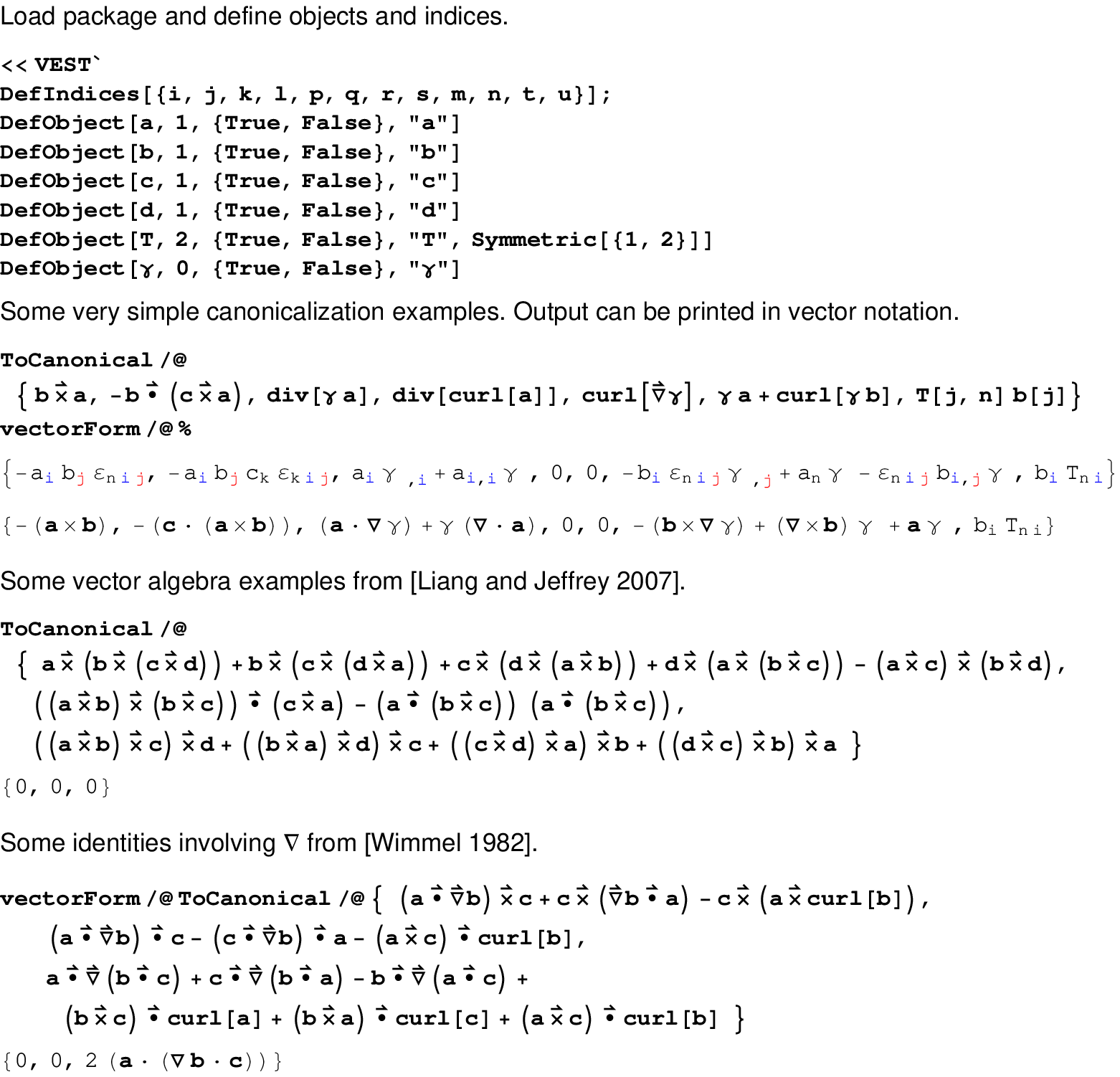}\caption{Various examples of the action of the \emph{VEST }function \texttt{ToCanonical}.\label{fig: ToCanonical}}

\par\end{centering}

\end{figure}
In Figure~\ref{fig: ToCanonical}, we give examples of the operation
of \texttt{ToCanonical} on various vector expressions. Note that input
and output can be in standard vector notation, indexed notation, or
a mix of both. Several vector algebra examples have been taken from
Refs.~\cite{Liang:2007p10618,Cunningham,Patterson,Stoutemyer}, with
some examples involving gradients taken from Ref.~\cite{Wimmel:1982p10623}.

\section{Simplification through Levi-Civita expansions\label{sec:Simplification-through-Levi-Civi}}

While adequate for simpler problems, the canonicalization of
individual terms as carried out by \texttt{ToCanonical} does not recognize
certain types of multi-term vector identities. In many cases, these
can lead to substantial simplification of large expressions. We have
found very few instances of this type of identity given in previous
literature; two examples are 
\begin{equation}
\bm{d}\left(\bm{a}\cdot\bm{b}\times\bm{c}\right)-\bm{a}\left(\bm{b}\cdot\bm{c}\times\bm{d}\right)+\bm{b}\left(\bm{c}\cdot\bm{d}\times\bm{a}\right)-\bm{c}\left(\bm{d}\cdot\bm{a}\times\bm{b}\right)=0,\label{eq:Simplest multi-term}
\end{equation}
which is relatively well known, and 
\begin{equation}
\bm{d}\left(\bm{a}\cdot\bm{b}\times\bm{c}\right)-\left(\bm{c}\cdot\bm{d}\right)\left(\bm{a}\times\bm{b}\right)-\left(\bm{a}\cdot\bm{d}\right)\left(\bm{b}\times\bm{c}\right)-\left(\bm{b}\cdot\bm{d}\right)\left(\bm{c}\times\bm{a}\right)=0,\label{eq:Simplest multi-term 2}
\end{equation}
which is given in \cite{Liang:2007p10618}. There are in fact whole
families of similar relations, including those involving gradient
tensors and more than four monomials. 

In essence, the ability to simplify expressions using these types of multi-term
vector identities requires two somewhat separate operations. Firstly,
we need to be able to generate vector identities that involve any given
monomial in the polynomial expression we wish to simplify. 
These identities can be used to construct
substitution rules for each term. Secondly, we require a way to sift through 
different combinations of these substitutions in order to arrive at the shortest
possible manifestation of the expression. 

In this
section we describe the \emph{VEST }function \texttt{FullSimplifyVectorForm},
which simplifies vector polynomials. \texttt{FullSimplifyVectorForm}
handles both of the aforementioned operations, \emph{deriving} vector
identities for each term before applying these in an attempt to find
the simplest form of an expression. We have found the function
to be very useful when carrying out large calculations, in many cases
reducing the size of expressions by more than a factor of two.

\subsection{Derivation of multi-term identities through Levi-Civita expansions\label{sub:Derivation-of-multi-term}}

Vector identities such as Eq.~(\ref{eq:Simplest multi-term}) can
be derived systematically using properties of the Levi-Civita symbol.
A simple technique used in \texttt{FullSimplifyVectorForm} is based
on the identity 
\begin{equation}
\frac{1}{2}\varepsilon_{irs}\varepsilon_{jrs}=\delta_{ij}.\label{eq:simple leviC identity}
\end{equation}
Starting from a single monomial, one substitutes Eq.~(\ref{eq:simple leviC identity})
for a chosen index or indices 
(using $a_i \rightarrow \delta_{ij}a_j=\varepsilon_{irs}\varepsilon_{jrs}a_j/2$), 
obtaining an equivalent term involving
more Levi-Civita symbols. Expanding pairs of Levi-Civita symbols in
various orders using Eq.~(\ref{eq:LeviC expansion}) will then sometimes
generate a non-trivially equivalent form, essentially a vector identity
involving the original monomial. 

We illustrate this process with a simple example that generates Eqs.~(\ref{eq:Simplest multi-term})
and (\ref{eq:Simplest multi-term 2}) above. Starting with $\bm{d}\left(\bm{a}\cdot\bm{b}\times\bm{c}\right)$
and inserting Eq.~(\ref{eq:simple leviC identity}) into the free
index, one obtains
\begin{equation}
a_{i}b_{j}c_{k}d_{n}\varepsilon_{ijk}=\frac{1}{2}a_{i}b_{j}c_{k}d_{l}\varepsilon_{ijk}\varepsilon_{lrs}\varepsilon_{nrs}.\label{eq:FSVF example}
\end{equation}
There are now three ways to expand pairs of Levi-Civita symbols using
Eq.~(\ref{eq:LeviC expansion}); expanding $\varepsilon_{ijk}\varepsilon_{nrs}$,
we are led to Eq.~(\ref{eq:Simplest multi-term}), while expanding
$\varepsilon_{ijk}\varepsilon_{lrs}$ gives Eq.~(\ref{eq:Simplest multi-term 2}).
Of course, an expansion of $\varepsilon_{lrs}\varepsilon_{nrs}$ will
simply generate the original term. As another example, the identities
\begin{align}
 & \left(\bm{a}\times\bm{c}\right)\cdot\nabla\bm{b}-\nabla\bm{b}\cdot\left(\bm{a}\times\bm{c}\right)-\bm{a}\left(\bm{c}\cdot\nabla\times\bm{b}\right)+\bm{c}\left(\bm{a}\cdot\nabla\times\bm{b}\right)=0,\nonumber \\
 & \left(-\bm{a}\cdot\bm{c}\right)\nabla\times\bm{b}+\bm{a}\times\left(\bm{c}\cdot\nabla\bm{b}\right)-\bm{a}\times\left(\nabla\bm{b}\cdot\bm{c}\right)+\bm{c}\left(\bm{a}\cdot\nabla\times\bm{b}\right)=0,\label{eq:Multi-term grad}
\end{align}
are generated from the monomial $\bm{c}\left(\bm{a}\cdot\nabla\times\bm{b}\right)$.
These identities {[}Eq.~(\ref{eq:Multi-term grad}){]} are of course
strongly related to those in Eqs.~(\ref{eq:Simplest multi-term})
and (\ref{eq:Simplest multi-term 2}), although in vector notation
the exact correspondence may not be immediately obvious. 

Finally, we note that the same process can be applied using more than
one pair of Levi-Civita symbols, replacing multiple indices in a monomial.
This will generate both a greater number of identities (due to the
large number of possible expansion orders) and larger individual identities
(since more symbols are involved) at the expense of computation time.

\subsection{Simplification of large vector expressions\label{sub:FullSimplifyVectorForm}}

We now describe the algorithm implemented in \texttt{FullSimplifyVectorForm}.
The general technique is to find equivalent forms (vector identities)
for each term and subsequently use these to bring the polynomial to
its shortest form. The method is iterative, continuing until the expression
is unchanged since the previous iteration. An outline of the procedure goes as follows:
\begin{enumerate}
\item Ensure expression is in canonical form.
\item For each monomial in the expansion, perform the procedure detailed
in Sec.~\ref{sub:Derivation-of-multi-term}. The identities
generated are stored as substitution rules so as to facilitate the subsequent search
for the shortest form. If desired, this step can be performed with
multiple pairs of Levi-Civita symbols, generating more substitution
rules at the expense of computational speed.
\item Consider the set of rules for each of the $n$ monomials generated in the 
previous step, denoting the rule set for monomial $i$ by $\mathcal{R}_i$
($\mathcal{R}_i$ includes the trivial substitution of a monomial into itself). It is desirable
to search through the entire set $\mathcal{R}_1 \times \ldots \times\mathcal{R}_n$ rather than
substituting just individual rules,
since rules containing similar terms can sometimes cancel when 
substituted concurrently.  Since the number of elements in $\mathcal{R}_1 \times \ldots \times \mathcal{R}_n$
can be very large (the product of the lengths of the $\mathcal{R}_i$),
\texttt{FullSimplifyVectorForm} reduces the size of this set by choosing 
only those elements that contain rules that have the possibility of canceling in a concurrent substitution. 
It then searches through all elements of this reduced set,
recording the length of the expression obtained for the 
substitutions given by each element. 

\item Choose the shortest expression obtained in step 3 and:

\begin{enumerate}
\item Return to step 1 if the expression has changed since the previous
iteration. Store rules found in previous step to save computation
time.
\item Return result if expression is unchanged since the previous iteration.
\end{enumerate}
\end{enumerate}
\begin{figure}
\begin{centering}
\includegraphics[width=1\textwidth]{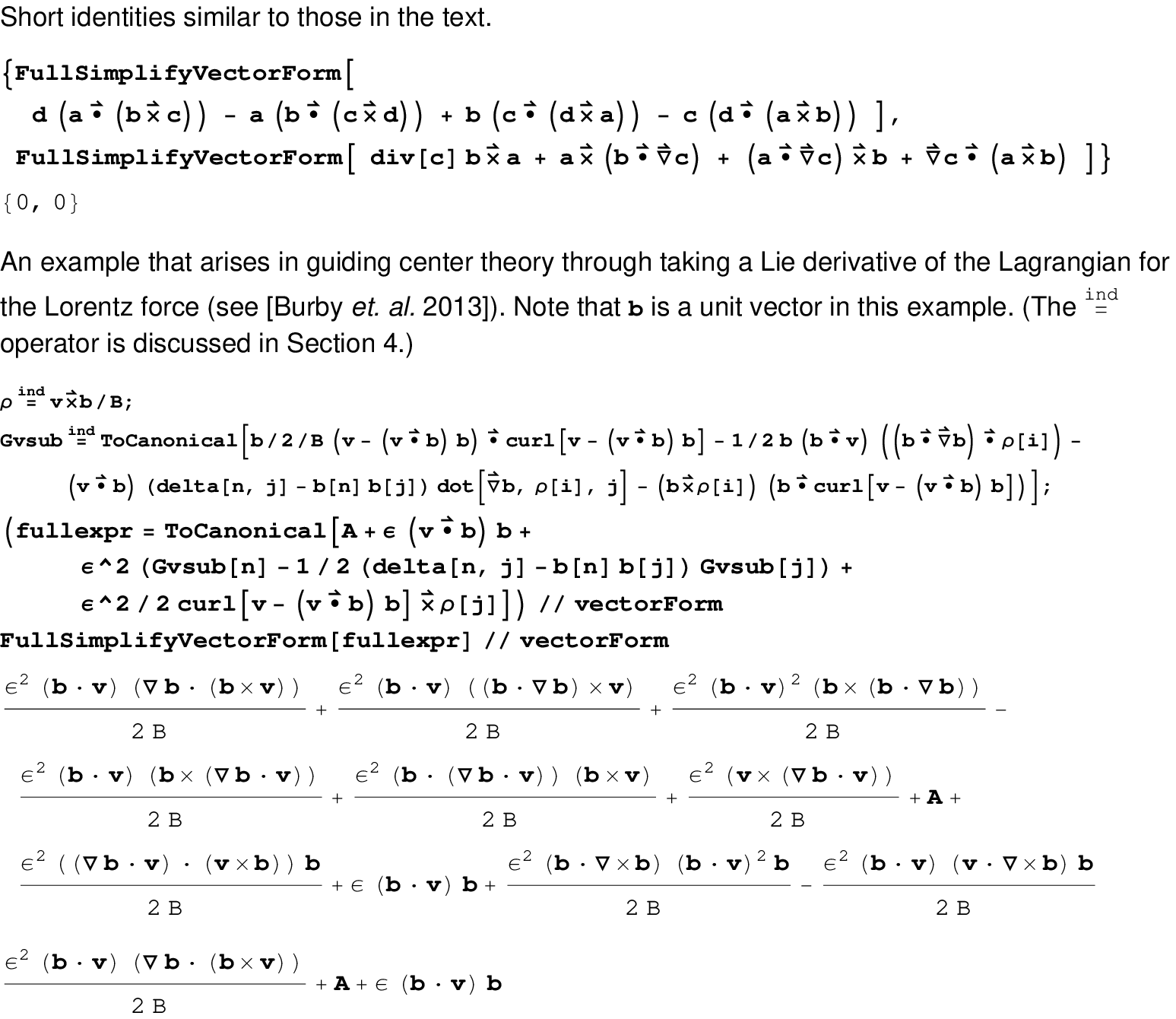}\caption{Some example applications of the \emph{VEST }function \texttt{FullSimplifyVectorForm}.
\label{fig:FullSimplifyVectorForm} }

\par\end{centering}

\end{figure}
In Figure~\ref{fig:FullSimplifyVectorForm} we give several examples
of the operation of \texttt{FullSimplifyVectorForm}. Due to the large
number of expansions to explore, the function can be relatively processor
intensive (especially if multiple pairs of Levi-Civita tensors are
used). The first two examples given in Figure~\ref{fig:FullSimplifyVectorForm}
take approximately 2 seconds to compute, while the final one requires
approximately 10 seconds. Memory use is not an issue for reasonable
expressions; we have simplified expressions of up to 600 monomials
without using more than roughly 100MB of memory.

\subsection{More general methods for automatic generation of vector identities
\label{sub:More-general-methods}}

While the method currently implemented in \texttt{FullSimplifyVectorForm}
(described above in Sec.~\ref{sub:Derivation-of-multi-term}) 
can easily and reliably generate many vector identities that  have
not appeared in previous literature, there are certain more complicated
identities that cannot be recognized. Specifically, any non-trivial
polynomial identity with no Levi-Civita symbols (\emph{i.e.,} involving \emph{only}
dot products) will not be identified by the above technique. 
If only one pair of Levi-Civita symbols is expanded at step 2 in the procedure 
this property is evident; insertion of the Levi-Civita pair into an expression without 
a Levi-Civita symbol leaves only one possibility of expansion ($\varepsilon_{irs}\varepsilon_{jrs}\rightarrow2\delta_{ij}$), 
trivially reproducing the original form. 
Interestingly, it seems that it is also not possible to produce identities lacking a Levi-Civita symbol when multiple
pairs are expanded in all possible combinations, although we have no proof that this is always true. These identities
that involve only dot products necessarily include more terms and larger
monomials (see below) than identities with Levi-Civita symbols, meaning they
are only required for simplification when very large expressions
(or expressions with multiple derivatives) are encountered. To give an example, 
with up to two derivatives and without involving unit
vector properties, the shortest such identity we have derived is given by
\begin{align}
0=&\left(\bm{a}^{2}\bm{c}^{2}-\left(\bm{a}\cdot\bm{c}\right)^{2}\right)\nabla\cdot\bm{b}^{2}-\left(\bm{a}^{2}\bm{c}^{2}-\left(\bm{a}\cdot\bm{c}\right)^{2}\right)b_{i,j}b_{j,i}+2\,\bm{a}^{2}\left(\bm{c}\cdot\nabla\bm{b}\cdot\nabla\bm{b}\cdot\bm{c}\right)\nonumber\\
+&\:2\,\bm{c}^{2}\left(\bm{a}\cdot\nabla\bm{b}\cdot\nabla\bm{b}\cdot\bm{a}\right)-2\,\left(\bm{a}\cdot\bm{c}\right)\left(\bm{a}\cdot\nabla\bm{b}\cdot\nabla\bm{b}\cdot\bm{c}+\bm{c}\cdot\nabla\bm{b}\cdot\nabla\bm{b}\cdot\bm{a}\right)\nonumber\\
+&\:2\,\left(\bm{a}\cdot\bm{c}\right)\nabla\cdot\bm{b}\left(\bm{a}\cdot\nabla\bm{b}\cdot\bm{c}+\bm{c}\cdot\nabla\bm{b}\cdot\bm{a}\right)+2\left(\bm{a}\cdot\nabla\bm{b}\cdot\bm{a}\right)\left(\bm{c}\cdot\nabla\bm{b}\cdot\bm{c}\right)\nonumber\\
-&\:2\left(\bm{a}\cdot\nabla\bm{b}\cdot\bm{c}\right)\left(\bm{c}\cdot\nabla\bm{b}\cdot\bm{a}\right)-2\,\nabla\cdot\bm{b}\left(\bm{c}^{2}\left(\bm{a}\cdot\nabla\bm{b}\cdot\bm{a}\right)+\bm{a}^{2}\left(\bm{c}\cdot\nabla\bm{b}\cdot\bm{c}\right)\right)\label{eq:multiterm vector ID}
\end{align}
for general vectors $\bm{a}$, $\bm{b}$ and $\bm{c}$. For this collection of objects 
(two of each $\bm{a}$, $\nabla\bm{b}$ and $\bm{c}$), there also exists a  slightly longer, 
very similar identity of 17 monomials. 
When unit vectors and/or more derivative tensors are included,
much shorter identities exist; for instance 
\begin{equation}
-b_{i,j}b_{j,k}b_{k,i}+\frac{3}{2}b_{i,j}b_{j,i}\nabla\cdot\bm{b}-\frac{1}{2}\left(\nabla\cdot\bm{b}\right)^{3}=0\label{eq:Littlejohn b identity}
\end{equation}
is given in Ref.~\cite{LittlejohnBID} for a unit vector $\bm{b}$ and 
can also be derived through the  method detailed in this
section\footnote{Eq.~\eqref{eq:Littlejohn b identity}
was derived in Ref.~\cite{LittlejohnBID} by noticing that it is nothing
but the Cayley-Hamilton theorem for matrix $b_{i,j}$ satisfying $\det \left(b_{i,j}\right)=0$. 
This method cannot be generalized to obtain other multi-term identities, but 
an interesting point is that the Cayley-Hamilton theorem is a simple consequence of the anti-symmetrization
procedure detailed in this section. Eq.~\eqref{eq:Littlejohn b identity} is the sole vector identity we have 
found in previous literature that cannot be recognized by the current 
version of \texttt{FullSimplifyVectorForm}}.
As another example, Eq.~\eqref{eq:multiterm vector ID} reduces to 8
terms when $\bm{a}$ or $\bm{c}$ is set to $\bm{b}$ and this is set as a
unit vector. This identity turned out to provide an important and non-trivial simplification
in our work described in Ref.~\cite{GCpaper}.

In this section we describe a very general technique for deriving 
vector identities that is not limited to identities that involve the cross-product. 
The method encompasses relations such as Eqs.~\eqref{eq:multiterm vector ID} and 
\eqref{eq:Littlejohn b identity}, as well identities with cross-products, 
\emph{e.g.,} Eqs.~\eqref{eq:Simplest multi-term} and \eqref{eq:Multi-term grad}.
While not implemented in the current version of \texttt{FullSimplifyVectorForm}, we 
will provide this functionality in a future release.
The overall approach is based on the idea that anti-symmetrization
of an $n$-dimensional tensor over $n+1$ indices will automatically
give a tensor polynomial that is identically zero. 
Such a polynomial
is necessarily of relatively high rank, so the construction of interesting
identities entails contracting over various pairs of indices. Note
that almost all such contractions trivially canonicalize to zero
and finding identities in this way by hand would be a very arduous
task. The method is essentially an 
application of Lovelock's "dimensionally dependent identities" of the 
Riemann tensor \cite{Lovelock:1970p10628} to tensor products of vectors and their gradients.

To be more precise, consider a general tensor $\mathcal{T}_{i_{1}\ldots i_{k}}$,
where the $i_{k}$ represent an arbitrary number of indices (note
that we work in 3-D Euclidean space with all lower indices). For the
cases we will consider $\mathcal{T}_{i_{1}\ldots i_{k}}$ will be
the product of vector objects, \emph{e.g., }$a_{i}b_{j}c_{k}d_{l}\varepsilon_{rsq}$.
Representing anti-symmetrization by $[\,]$ around relevant indices,
the identity
\begin{equation}
\mathcal{T}_{i_{1}\ldots i_{j-1}i_{j+1}\ldots i_{k}\left[i_{j}\right.}\delta_{a_{1}}^{b_{1}}\delta_{a_{2}}^{b_{2}}\delta_{\left.a_{3}\right]}^{b_{3}}=0\label{eq:Asym Tensor ID}
\end{equation}
must hold for for all $1\leq j\leq k$, since the tensor is anti-symmetric
over four indices in three dimensions. In Eq.~(\ref{eq:Asym Tensor ID})
$\delta_{i}^{j}$ is simply the standard Kronecker delta $\delta_{ij}$,
we write with an ``up'' index to more clearly show the anti-symmetrization, note the non-standard use of this notation.
One can also anti-symmetrize over more indices of $\mathcal{T}$ (and
fewer $\delta_{ij}$) if desired, but the resulting identity will
involve only the relevant anti-symmetric part of $\mathcal{T}$. Of
course, in the case where $\mathcal{T}$ is already anti-symmetric
in some set of indices (in our case due to $\varepsilon_{ijk}$),
an identity involving the entirety of $\mathcal{T}$ with fewer $\delta_{ij}$
can be obtained by anti-symmetrizing over these indices. For example,
with $\mathcal{T}$ anti-symmetric over $\left\{ i_{1},i_{2},i_{3}\right\} $,
\begin{equation}
\mathcal{T}_{i_{4}\ldots i_{k}\left[i_{1}i_{2}i_{3}\right.}\delta_{\left.a_{1}\right]}^{b_{1}}=0.\label{eq:Asym Ten ID 2}
\end{equation}
This insight explains why vector identities that involve $\varepsilon_{ijk}$
can be so much simpler than those that do not; any tensor with anti-symmetry
will naturally have identities with fewer terms than those without
anti-symmetry, since non-trivial identities in the form of Eq.~(\ref{eq:Asym Tensor ID})
can be constructed with fewer indices. 

The ideas of the previous paragraph can be used to automatically generate
vector identities from a given set of vector objects. Although certainly
not the most efficient method, a simple algorithm goes as follows:
\begin{enumerate}
\item For a given vector monomial, consider the tensor obtained by removing
all contractions between dummy indices \emph{e.g., }for $a_{i}a_{j}b_{i}b_{l}b_{j,l}b_{k,k}$
this is $a_{i}a_{j}b_{k}b_{l}b_{r,s}b_{p,q}$.
\item Choose an index over which to anti-symmetrize and form the polynomial
given by Eq.~(\ref{eq:Asym Tensor ID}). If one of the objects from
step 1 is $\varepsilon_{ijk}$, construct Eq.~(\ref{eq:Asym Ten ID 2})
instead, antisymmetrizing over the indices of $\varepsilon_{ijk}$
(the reason for this is simply to generate shorter identities).
For instance, with the tensor example given in step 1, one could use
$a_{i}a_{j}b_{k}b_{l}b_{r,s}b_{p,[q}\delta_{a}^{b}\delta_{c}^{d}\delta_{e]}^{f}$.
\item Contract the polynomial between index pairs. Aside from
those contractions that are known \emph{a priori }to give identically
zero (see Ref.~\cite{edgar:659}) all possible contractions should be evaluated.
For example, $a_{i}a_{j}b_{k}b_{l}b_{r,s}b_{s,[q}\delta_{i}^{q}\delta_{r}^{j}\delta_{l]}^{k}$
is the scalar formed by contraction of the tensor given above between
the index pairs \\$\left\{ \left\{ 1,9\right\} \left\{ 2,12\right\} \left\{ 3,14\right\} \left\{ 4,13\right\} \left\{ 5,11\right\} \left\{ 6,7\right\} \left\{ 8,9\right\} \right\} $. If one wishes  to generate only those identities
 involving the original scalar monomial 
 (\emph{i.e.,} $a_{i}a_{j}b_{i}b_{l}b_{j,l}b_{k,k}$ in the running example), 
consider only the set of contractions that have a possibility of generating 
this.
\item Canonicalize the resulting list of scalar or vector expressions to
remove $\delta_{ij}$ and cancel relevant terms.
\end{enumerate}
We have applied this procedure
to various forms, systematically generating identities that involve
a given set of objects, both with and without the Levi-Civita symbol.
For instance, applying the method to the objects $\left\{ \bm{a},\bm{b},\bm{c},\bm{d},\varepsilon\right\} $
(\emph{i.e., }the tensor $a_{i}b_{j}c_{k}d_{l}\varepsilon_{rsq}$
at step 2) generates Eqs.~(\ref{eq:Simplest multi-term}) and (\ref{eq:Simplest multi-term 2}), 
Eq.~\eqref{eq:multiterm vector ID} is generated with 
$\left\{ \bm{a},\bm{a}, \bm{c},\bm{c},\nabla\bm{b},\nabla\bm{b}\right\} $,
and Eq.~(\ref{eq:Littlejohn b identity}) is generated with $\left\{ \bm{b},\bm{b},\nabla\bm{b},\nabla\bm{b},\nabla\bm{b}\right\} $
and a subsequent application of various unit vector identities. There
are of course many other similar relations that we have not listed here.

We note that the \emph{Invar} package \cite{Invar}
uses a similar anti-symmetrization based method as part of its algorithm to
 generate scalar invariants of the Riemann tensor.

\section{Additional \emph{VEST} functionality\label{sec:Additional-VEST-functionalit}}

In addition to the functions described in Sections~\ref{sec:Index-notation-as}
and \ref{sec:Simplification-through-Levi-Civi}, \emph{VEST }contains
several other features than can be very useful when carrying out large
calculations. In this section we briefly describe some of this functionality.

\subsection{Intuitive and user friendly input and output}

While very precise and straightforward to interpret, index notation
can be inconvenient for the user, since expressions often look jumbled
and confusing. As illustrated in Figs.~\ref{fig: ToCanonical} and
\ref{fig:FullSimplifyVectorForm}, \emph{VEST} includes several
features to facilitate user input and output, both in index and vector
notation. Expressions can be input in standard vector notation omitting
indices (e.g., $\mathtt{curl[a]}$), full index notation, or a mix
of both (e.g., $\mathtt{div[b[i]v[i]b[j]]+a\centerdot\left(T[i,j]b[j]\right)}$).
This allows for fast and reliable user input with the ability to represent
more complex expressions where vector notation becomes ambiguous.
In addition to coloring dummy index pairs so contractions are more
immediately obvious, the function \texttt{vectorForm} prints expressions
using vector notation where possible (up to first order derivatives),
see Figs.~\ref{fig: ToCanonical} and \ref{fig:FullSimplifyVectorForm}.

\subsection{Checking expression equality}

The function \texttt{CheckTensorZero} provides a very reliable check
of whether an expression is identically zero. This is useful both
for when one is not confident that \texttt{FullSimplifyVectorForm}
has reached the shortest possible form and for rapid verification
of results. The function works in a very straightforward way by expanding
an expression into Cartesian co-ordinates, which amounts to explicitly
evaluating all sums over dummy indices. If so desired, the user can
specify particular forms for some (or all) objects in an expression.
This is useful both for when non-trivial relationships exist between
different objects (e.g., $\bm{B}=\nabla\times\bm{A}$) and in the
expansion of very large expressions where memory use becomes an issue
and a numerical check is necessary. \texttt{CheckTensorZero} also
includes functionality to search through an expression for subsets
that are zero, which has occasionally proved useful for simplification
purposes.

\subsection{Substitutions}

A very common application of a computer algebra package is the substitution
of some explicit expression into a given form, \emph{i.e.,} given a specific
$a$, calculate $f\left(a\right)$. While this is a very simple process
for standard algebraic expressions, the task becomes more awkward when the
substitution involves indexed expressions. To illustrate this, consider
as a basic example the evaluation of 
\begin{equation}
a_{i}a_{j,k}b_{j}b_{k},\;\mathrm{with}\; a_{i}=b_{j}d_{j}d_{i}.\label{eq:ind sub}
\end{equation}
There are two issues that arise if one attempts a naive substitution
of $a_{i}$; first, the free index of $a_{i}=b_{j}d_{j}d_{i}$ must
be replaced with the correct indices in $a_{i}a_{j,k}b_{j}b_{k}$,
and second, one must ensure that dummy indices in the substituted
$a_{i}$ do not conflict with those in $a_{i}a_{j,k}b_{j}b_{k}$.
While these issues are in principle not complicated, forcing the user
to keep track of all indices would be a particularly inconvenient
characteristic that would significantly reduce the utility of an index
notation based package. 

In \emph{VEST }substitution of arbitrary expressions is handled
through a new assignment operator $\mathtt{\overset{ind}{=}}$, which
automatically manages assignment of free indices and ensures dummies
do not overlap. Rather than simply assigning an expression to the
left hand side, $\mathtt{\overset{ind}{=}}$ assigns a call to the
function \texttt{FindDummies}, which is used in step 2 of \texttt{ToCanonical}
(see Sec.~\ref{sub:Canonicalization}) and can generate a new set
of indices at every call. After assignment with $\mathtt{\overset{ind}{=}}$,
an object can be used in exactly the same way as a standard indexed
object without the user having to worry about its underlying structure.
To illustrate with the example of Eq.~(\ref{eq:ind sub}), after
assigning $\mathtt{a\overset{ind}{=}b[j]d[j]d[i]}$, evaluating $ $$a_{i}a_{j,k}b_{j}b_{k}$
in the standard way will generate a valid indexed expression.

\subsection{Unit vectors and user defined rules}

Expressions involving unit vectors arise often in certain types of
calculations, including the guiding center calculation that we have
discussed regularly throughout the manuscript. As several examples
earlier in the text have illustrated (\emph{e.g.,} Sec.~\ref{sub:More-general-methods}),
unit vector identities can provide very substantial simplifications
and it is important to make provision for these. Representing an arbitrary
unit vector by $b_{i}$, \emph{VEST} automatically generates identities
by differentiating $b_{i}b_{i}=1$ up to a user-specified order
and applies these rules as part of \texttt{ToCanonical}. In addition,
unit vectors are automatically accounted for in \texttt{FullSimplifyVectorForm}
by multiplying each term by $b_{i}b_{i}$ before each Levi-Civita
expansion (see Sec.~\ref{sub:Derivation-of-multi-term}), and in
\texttt{CheckTensorZero} by setting $b_{3}=\sqrt{1-b_{1}^{2}-b_{2}^{2}}$
for any unit vectors in an expression.

A related feature is the ability for the user to define rules that
are applied as part of \texttt{ToCanonical}. This is very useful both
when non-trivial relationships between objects need to be identified
(e.g., $\nabla\cdot\left(B\bm{b}\right)=0$) and when working with
expressions that involve non-trivial scalar expressions in the denominator.

\section{Conclusion}

In this paper, we have presented a new \emph{Mathematica} package,
\emph{VEST} (Vector Einstein Summation Tools), for performing abstract vector calculus computations.
Routines for reduction to standard form and simplification are based on representation
of vector polynomials using index notation with the Einstein
summation convention. The utility of the package has been illustrated
through multiple calculations of the single particle guiding center
Lagrangian in our companion paper \cite{GCpaper}, which is usually
a long and arduous process requiring months of tedious algebra. 

The \texttt{ToCanonical} function in \emph{VEST} encompasses almost
all previously published vector identities \cite{NRL,Liang:2007p10618,Wimmel:1982p10623}
and provides a very thorough reduction for most expressions. For larger
polynomials, more comprehensive simplification capabilities are provided
by the function \texttt{FullSimplifyVectorForm}, which uses expansions
of Levi-Civita symbols to derive multi-term vector identities. \texttt{FullSimplifyVectorForm}
has proven to be very useful in practice, often reducing the length
of expressions dramatically in ways that would be very difficult to find by hand.
We note that in previous literature we have found only a handful of examples
of the type of identity derived by \texttt{FullSimplifyVectorForm}. 

In addition to the method of Levi-Civita expansions, we have illustrated
a more general technique based on anti-symmetrization that can be
used to derive very general non-trivial identities for a given set of objects.
As future work, we hope to implement some variant of this technique
into \texttt{FullSimplifyVectorForm}, which would allow the simplification of expressions that 
do not contain cross-products. To improve computational 
efficiency, rules can be pre-calculated and stored in look-up tables, a method that is 
used in the \emph{Invar} package \cite{Invar} to simplify polynomials of the 
Riemann tensor. While the generation of basic identities will be simpler than
those for the Riemann tensor (due to its complex symmetries), the substitution of known identities
will be complicated somewhat by allowing multiple different objects,
including those of different rank. Another significant complication
will be the inclusion of unit vectors, the properties of which can significantly change
identities in non-trivial ways. 

\emph{VEST} has been designed to be very user-friendly, with intuitive
and simple input and output. We invite the reader to try out the package,
which can be found along with a comprehensive tutorial from the CPC program 
library.

\section*{Acknowledgements}

This research is supported by U.S.~DOE (DE-AC02-09CH11466).


\begin{thebibliography}{10}
\expandafter\ifx\csname url\endcsname\relax
  \def\url#1{\texttt{#1}}\fi
\expandafter\ifx\csname urlprefix\endcsname\relax\def\urlprefix{URL }\fi
\expandafter\ifx\csname href\endcsname\relax
  \def\href#1#2{#2} \def\path#1{#1}\fi

\bibitem{GCpaper}
J.~W. Burby, J.~Squire, H.~Qin, Automation of the guiding center expansion,
  Physics of Plasmas 20~(7) (2013) 072105.

\bibitem{Mathematica}
Mathematica, Version 9.0, Wolfram Research Inc., Champaign, Illinois, 2012.

\bibitem{Eastwood1991121}
J.~W. Eastwood, {ORTHOVEC}: version 2 of the {REDUCE} program for 3-{D} vector
  analysis in orthogonal curvilinear coordinates, Computer Physics
  Communications 64~(1) (1991) 121--122.

\bibitem{VecCalc}
P.~B. Yasskin, A.~Belmonte, CalcLabs with Maple commands for multivariable
  calculus, fourth edition, Brooks/Cole, Cengage Learning, Belmont, CA, 2010.

\bibitem{Vectan}
B.~Fiedler, Vectan 1.1, manual Math. Inst., Univ. Leipzig. (1997).

\bibitem{WIRTH:1979p10629}
M.~C. Wirth, Symbolic vector and dyadic analysis, Siam J. Comput. 8~(3) (1979)
  306.

\bibitem{Liang:2007p10618}
S.~Liang, D.~J. Jeffrey, Rule-based simplification in vector-product spaces,
  Towards Mechanized Mathematical Assistants, Lecture Notes in Artificial
  Intelligence (2007) 116--127.

\bibitem{Qin:1999p5832}
H.~Qin, W.~Tang, G.~Rewoldt, Symbolic vector analysis in plasma physics,
  Computer physics communications 116 (1999) 107--120.

\bibitem{Maple}
Maple {W}aterloo {S}oftware inc., see http://www.maplesoft.com.

\bibitem{CARY:1981p8578}
J.~R. Cary, Lie transform perturbation theory for {H}amiltonian systems, Phys.
  Rep. 79~(2) (1981) 129--159.

\bibitem{Cary:2009p8641}
J.~R. Cary, A.~J. Brizard, Hamiltonian theory of guiding-center motion, Rev.
  Mod. Phys. 81~(2) (2009) 693--738.

\bibitem{Littlejohn:1983p5809}
R.~G. Littlejohn, Variational principles of guiding centre motion, Journal of
  Plasma Physics 29~(1).

\bibitem{NRL}
J.~D. Huba, {NRL} {P}lasma {F}ormulary (2009).

\bibitem{MacCallum:2002p10643}
M.~A.~H. MacCallum, Computer algebra in general relativity, International
  Journal of Modern Physics A 17~(20) (2002) 2707--2710.

\bibitem{Wang13}
Y.~Wang, Math{GR}: a tensor and {GR} computation package to keep it simple,
  arXiv: abs/1306.1295.

\bibitem{MathTensor}
L.~Parker, S.~Christensen, MathTensor: {A} system for doing tensor analysis by
  computer, Addison-Wesley, Reading, MA, 1994.

\bibitem{Peeters2007550}
K.~Peeters, Cadabra: a field-theory motivated symbolic computer algebra system,
  Computer Physics Communications 176~(8) (2007) 550--558.

\bibitem{xAct}
J.~M. Mart\'{i}n-Garc\'{i}a, \href{http://www.xact.es/}{Efficient tensor
  computer algebra} (2002--2012).
\newline\urlprefix\url{http://www.xact.es/}

\bibitem{Tensorial}
J.~F. Gouyet, R.~Cabrera, D.~Park,
  \href{http://www.jfgouyet.fr/tcm/tcm.html}{Tensorial 4.0 and
  {TC}ontinuum{M}echanics 2.0} (2008).
\newline\urlprefix\url{http://www.jfgouyet.fr/tcm/tcm.html}

\bibitem{Stoutemyer}
D.~Stoutemyer, Symbolic computer vector analysis, Computers and Mathematics
  with Applications 5 (1979) 1--9.

\bibitem{MartinGarcia:2008p10621}
J.~M. Mart\'{i}n-Garc\'{i}a, \emph{xPerm}: fast index canonicalization for
  tensor computer algebra, Computer physics communications 179~(8) (2008)
  597--603.

\bibitem{Manssur:2002p10644}
L.~R.~U. Manssur, R.~Portugal, B.~Svaiter, Group-theoretic approach for
  symbolic tensor manipulation, International Journal of Modern Physics C
  13~(07) (2002) 859--879.

\bibitem{Cunningham}
J.~Cunningham, Vectors, Heinemann Educational Books Ltd, London, 1969.

\bibitem{Patterson}
E.~Patterson, Solving problems in vector algebra, Oliver and Boyd Ltd,
  Edinburgh-London, 1968.

\bibitem{Wimmel:1982p10623}
H.~K. Wimmel, Extended standard vector analysis with applications to plasma
  physics, European Journal of Physics 3~(4) (1982) 223.

\bibitem{LittlejohnBID}
R.~G. Littlejohn, Geometry and guiding center motion, Contemporary mathematics
  28 (1984) 151.

\bibitem{Lovelock:1970p10628}
D.~Lovelock, Dimensionally dependent identities, Math. Proc. Camb. Phil. Soc.
  68~(2) (1970) 345.

\bibitem{edgar:659}
S.~B. Edgar, A.~Hoglund, Dimensionally dependent tensor identities by double
  antisymmetrization, Journal of Mathematical Physics 43~(1) (2002) 659--677.

\bibitem{Invar}
J.~M. Mart\'{i}n-Garc\'{i}a, R.~Portugal, L.~Manssur, The invar tensor package,
  Computer Physics Communications 177~(8) (2007) 640 -- 648.

\end{thebibliography}

\end{document}